# Topological Properties of Web Services Similarity Networks

Chantal Cherifi[1], Vincent Labatut[2] and Jean-François Santucci[1],

[1] University of Corsica, UMR CNRS, SPE Laboratory, France

[2] Galatasaray University, Computer Science Department, Ortaköy Istanbul, Turkey

**Abstract.** The number of publicly available Web services (WS) is continuously growing. To perform efficient WS discovery, it is desirable to organize the WS space. Works in this direction propose to group WS according to certain shared properties. Such groups commonly called communities are based either on similarity or on interaction between WS. In this paper we focus on the former, and propose a new network-based approach to extract communities from a WS collection. This process is three-stepped: first we define several similarity functions able to compare WS operations, second we use them to build so-called similarity networks, and third we identify communities under the form of specific structures in these networks. We apply our method on a collection of real-world WS and comment the resulting communities. Finally, we additionally provide an analysis and an interpretation of our similarity networks with a complex networks perspective.

## 1  Introduction

A Web Service (WS) is an autonomous software component which can be published, discovered and invoked for remote use. When a provider creates a new WS, he describes it using a WS description language. A WS description file is comparable to an interface defined in the context of object-oriented programming: it lists the operations implemented by the WS. Currently, production WS use syntactic descriptions expressed with the WS description language (WSDL), which is a W3C (World Wide Web Consortium) recommendation. Such descriptions basically contain the names of the operations and their parameters names and data types, plus some lower level information regarding the network access to the WS.

To make a WS available to consumers, the provider records it in a registry by supplying the appropriate information, including the WS description. When a consumer wants to use a WS, it queries the registry to find one that matches his needs and obtain its access point. But finding the right WS is not an easy task. Indeed the number of available WS is continuously growing. Furthermore WS are volatile; they often operate in a highly dynamic environment as providers remove, modify, or relocate them frequently. Hence, it may be of great help to organize the WS space in a meaningful manner, in order to facilitate WS discovery, composition and substitution. To that end, the WS classification process aims at grouping WS into categories usually called communities. Most works in WS classification have focused on two classification types: grouping WS according to their similarity (Medjahed & Bouguettya, 2005; Taher, Benslimane, Fauvet & Maamar, 2006; Benatallah, Dumas, Sheng & Ngu, 2002) or to their possible interaction (Dekar & Kheddouci, 2008; Sepehrifar, Zamanifar & Sepehrifar, 2009).

In this work we focus on the former, and propose a new, network-based approach to build communities. Our approach is three-stepped: first we use several similarity functions to compare every pair of operations in a given WS collection, second we build a so-called similarity network from these values, and third we extract communities from this network. In parallel, we perform a detailed analysis and interpretation of the similarity network topological properties.

The rest of the paper is organized as follows. In section 2, we review the existing approaches regarding WS communities definition. We also explain how these communities can be used practically. Indeed, although this point is out of the scope of this article, it should be noted the communities mined with our own method can be used for the same purposes. Section 3 provides key information about complex networks, their analysis and properties. In section 4, we introduce similarity networks and the method we use to build them. In section 5 we present and discuss our results, focusing on the comparison of the proposed similarity functions and on the similarity networks properties. Conclusions and perspectives are given in section 6.

## 2 Related Works

Existing works adopt either a top-down or a bottom-up approach to define communities. In the former, abstract communities are designed a priori, and WS are then defined in order to fit these categories. In the later, communities are mined from an existing WS collection. They also can be distinguished depending on the type of WS description they use (syntactic vs. semantic), and their definition of the concept of similarity.

The top-down approach used in (Medjahed & Bouguettya, 2005) consists in defining WS similarity relatively to their application domains. They introduced the concept of community ontology, which serves as a template for describing communities of WS operating in the same thematic area, independently from their functionalities. The authors use communities to tackle the problems of organizing, describing, and managing semantic WS. A community is view itself as a WS which is created, advertised, discovered, and invoked in the same way regular WS are.

Benatallah *et al*. define a community as a group of WS sharing a common functionality, although they can simultaneously have different non-functional properties, like different providers or QoS parameters (Benatallah, Dumas, Sheng & Ngu, 2002). A community is considered as a set of alternative WS represented by a set of generic operations. It is used in the context of discovery and composition. When some request targets a given functionality, the corresponding community is identified and the request is delegated to one of its members. The choice of the delegate is based on the parameters of the request, the characteristics of the members, the history of past executions and the status of ongoing executions.

Other authors use functionality-based similarity, but with a semantic web approach (Taher, Benslimane, Fauvet & Maamar, 2006). Each community is associated to a specific functionality, shared by all the WS it contains, and represented by an ontological concept. A community is then defined as a triple containing an abstract WS (community functionality described by some abstract operations), a set of concrete WS and a mapping module between the abstract operations and the concrete WS. A community is considered as a set of substitutable WS (relatively to the community functionality) and is advertised in a UDDI registry for discovery purpose.

Unlike the first three ones, this work and the following ones follow a bottom-up approach. Nayak *et al.* (Nayak & Lee, 2007) computed separately different similarities between WS on UDDI descriptions, WSDL terms and OWL-S terms using Jaccard coefficient. Similarities are then merged to obtain a so-called accumulative similarity. By proceeding likewise for all pairs in a WS collection, the authors build a similarity matrix, on which a clustering algorithm is applied to identify the communities.

Konduri *et al* (Konduri & Chan, 2008) use the interface similarity assessment method (Wu & Wu, 2005) to compute semantic similarity on operations and parameters name. This method compares several types of WS properties (common properties, special properties, WS interface and QoS) to get an overall similarity measure used to fill a similarity matrix. Like in the previous approach, communities are then identified by clustering. Each community is represented by a set of characteristic operations designed to be used for WS discovery.

Instead of using the community notion, Kona *et al*. developed a theory of substituability (Kona, Bansal, Simon, A.Mallya, Gupta & Hite, 2006). A semantical WS can be a substitute to another if it requires as many or less inputs and if it produces as many or more outputs. This holds both in terms of number of parameters, and ontological level of the concepts associated to the parameters. The theory was designed to build tools for automatically discovering and composing WS.

Our approach is bottom-up and relies on syntactic WS descriptions, although it can be easily extended to semantic ones. The main differences are the use of four original distinct similarity functions, each one with its own interpretation, and the fact the communities are identified as specific structures in a complex network. Moreover, we focus on the operation level, by opposition to the WS level. In all four functions, the similarity is defined in terms of functionalities.

## 3  Complex Network Properties

Complex networks are a specific class of graphs, characterized by a huge number of nodes and non trivial topological properties. Used in many different fields to model real-world systems (Costa, Oliveira, Travieso, Rodrigues, Boas, Antiqueira, Viana & Rocha, 2008), they have been intensively studied both theoretically and practically (Newman, 2003). Because of their complexity, specific tools are necessary to analyze and compare them. This is usually performed through the comparison of several well-known properties, supposed to summarize the essential of the network structure.

**Components Organization**

A *component* is a maximal connected sub graph, i.e. a set of interconnected nodes, all disconnected from the rest of the network. The component distribution and, more specifically, the size of the largest component are important network properties. The fact the network is split in several separated parts with various sizes is directly related to the modeled system effectiveness at doing its job. In some cases, a so-called giant component, whose size is far greater than the other components, is required for the system to work efficiently. For example, in a communication network like the Internet, the size of the largest component represents the largest fraction of the network within which communication is possible (Newman, 2003). Most real-world networks have a giant component. However, in some other cases, separated small parts are preferable. For instance, when epidemiologists model disease propagation in a population, they look for scattered networks.

**Average Distance**

The *distance* between two nodes is defined as the number of links in the shortest directed path connecting them. At the level of the whole network, this allows to process the *average distance* and the diameter. The former corresponds to the mean distance over all pairs of nodes (Newman, 2003). This notion is related to the *small world* property, observed when this distance is relatively small. The classic procedure to assess this property consists in comparing the average distance measured in some network of interest to the one estimated for an Erdős–Rényi (ER) network (Erdos & Renyi, 1959) containing the same numbers of nodes and links, since this random generative model is known to produce networks exhibiting the small world property (Newman, 2003). In terms of dynamic processes, the existence of shortcuts between nodes can be interpreted as propagation efficiency (Watts & Strogatz, 1998). Most real-world networks have the small world property.

**Transitivity**

Network *transitivity* (also called clustering) corresponds to the triangle density in the considered network, where a triangle is a structure of three completely connected nodes. It is measured by a transitivity coefficient, which is the ratio of existing to possible triangles in the network (Watts & Strogatz, 1998). The higher this coefficient, the more probable it is to observe a link between two nodes which are both connected to a third one. A real-world network is supposed to have a higher transitivity than the corresponding ER network by an order of magnitude corresponding to their number of nodes, meaning their nodes tend to form densely connected groups.

## 4 Similarity Networks

Networks constitute a convenient way to represent a collection of WS, allowing visualizing, analyzing and taking advantage of the relationships of similarity observed between them. Generally speaking, we define a *similarity network* as a graph whose nodes correspond to objects, and links indicate a certain similarity between the connected nodes. They can be considered as complex networks and some authors previously used this approach to model WS collections in other contexts than community identification (Liu, Liu & Chao, 2007; Oh, Lee & Kumara, 2008; Talantikite, Aissani & Boudjlida, 2009; Gekas & Fasli, 2008; Shiaa, Fladmark & Thiell, 2008; Kwon, Park, Lee & Lee, 2007; Hashemian & Mavaddat, 2005; Cherifi, Labatut & Santucci, 2010 -a; Cherifi, Labatut & Santucci, 2010 -b).

To build our similarity networks, we decided, as a first step, to focus on syntactically described operations. An operation is a part of a WS implementing a specific functionality. It is syntactically described by its name and its input and output parameters (names and data types). To represent a collection of WS descriptions under the form of a similarity network of operations, we first create a node to represent each operation in the collection. Then, a link is added between two nodes iff the corresponding operations present a certain similarity. In the resulting network, similar operations are connected and form graph components. We previously defined a community as a set of similar operations, so identifying them is straightforward here: each component simply corresponds to a community. Of course, the nature of the similarity relation is extremely important, and can be defined in various ways. In the following, we describe four similarity functions and explain how they can be interpreted and used.

**Similarity Functions**

In our case, a similarity function $f$ takes two operations $o_1$ and $o_2$ and quantifies their similarity. It can be either symmetrical ($f(o_1, o_2) = f(o_2, o_1)$) or asymmetrical, and it can output binary or real values. The four functions we defined to build our networks are based on previous works focusing on WS retrieval (Keller, Lara, Lausen, Polleres & Fensel, 2005; Küster & König-Ries, 2008). In these articles, the authors defined several matchmaking operators and used them to compare sets of ontological concepts. We selected four of these operators: match, partial match, excess match, relation match, and adapted them to our goal, which is the processing of a similarity value between two sets of parameters. We obtained four similarity functions we called *Full Similarity*, *Partial Similarity*, *Excess Similarity* and *Relation Similarity*; all of them with a binary output. These functions are defined in terms of set relations between the input and output parameter sets of the compared operations.

Let $I_i$, and $O_i$, be the sets of input and output parameters for operation $o_i$, respectively. Suppose we want to compare $o_1$ and $o_2$. *FullSim* is a symmetrical function stating both operations are fully similar iff 1) they provide exactly the same outputs ($O_1 = O_2$) and 2) they need overlapping inputs ($I_1 \cap I_2 \neq \emptyset$). *PartialSim* and *ExcessSim* are asymmetrical. With the former, $o_2$ is partially similar to $o_1$ iff 1) some $o_1$ outputs are missing in $o_2$ ($O_1 \supset O_2$) and 2) they need overlapping inputs ($I_1 \cap I_2 \neq \emptyset$). With the latter, $o_2$ is similar to $o_1$ with excess iff 1) $o_2$ provides all $o_1$ outputs plus additional ones ($O_1 \subset O_2$) and 2) $o_2$ needs only some of $o_1$ inputs ($I_1 \supseteq I_2$). The *RelationSim* function is symmetrical and states both operations have a relational similarity iff 1) they have exactly the same outputs ($O_1 = O_2$) and 2) they share no common input ($I_1 \cap I_2 = \emptyset$). These definitions are summarized in Table 1.

To determine the relations between two sets of parameters, one needs to be able to compare the parameters themselves. For this purpose, we selected the basic matching operator already used in our previous work (Cherifi, Labatut & Santucci, 2010 -a), consisting in comparing only the parameters names. Two parameters are said to be equal iff their names are the exact same strings. This is a very simple and rather naïve operator as it certainly leads to irrelevant matching and miss some relevant ones, from a semantic point of view. Nevertheless, for this first work, we chose to put our focus on building communities and studying their network properties, rather than trying more flexible matching operators.

To summarize this section: we defined four different similarity functions able to compare operations, all using the same simple matching operator to compare parameter names. Each function corresponding to a different definition of the concept of similarity, they will lead to different similarity networks when applied to a given WS collection.

**Table 1** *Similarity Functions Definitions*

| Similarity function | Sets Relations | Direction |
|---|---|---|
| *FullSim* | $(I_1 \cap I_2 \neq \emptyset) \wedge (O_1 = O_2)$ | *symetrical* |
| *PartialSim* | $(I_1 \cap I_2 \neq \emptyset) \wedge (O_1 \supset O_2)$ | *assymetrical* |
| *ExcessSim* | $(I_1 \supseteq I_2) \wedge (O_1 \subset O_2)$ | *assymetrical* |
| *RelationSim* | $(I_1 \cap I_2 = \emptyset) \wedge (O_1 = O_2)$ | *symetrical* |

**Interpretation**

We will now use an example to show how our similarity functions can be interpreted, and why they are relevant to compare operations. Consider the operations defined in Figure 1: $o_1$ is called `get_CITYNAMEbyZIP` and returns the city name corresponding to the specified zip code; $o_2$ is called `get_CITYNAMEbyZIPGEOGRAPHICALREGION` and returns the city name corresponding to the specified zip code and geographical region, $o_3$ is called `get_GEOGRAPHICALLOCATIONbyZIP` and returns the city name, longitude, latitude and altitude of the location corresponding to the specified zip code, $o_4$ is called `get_WEATHERbyZIP` and returns the weather report of the location corresponding to the specified zip code, $o_5$ is called `get_WEATHERbyCITYNAME` and returns the weather report of a city whose name has been specified, and $o_6$ is called `get_WEATHERWEATHERREPORTSUBSCRbyCITYNAME` and returns the weather report itself and a subscription form for a city whose name has been specified. $o_1$ and $o_2$ are fully similar, because they produce the same outputs and have common inputs. $o_1$ is partially similar to $o_3$, because it lacks some of $o_3$ outputs and they have common inputs. $o_6$ is similar to $o_5$ with excess, because it produces all of $o_5$ outputs and more, and they have common inputs. Finally, $o_4$ and $o_5$ are relationally similar, because their outputs are the same and they have no common input.

The meaning of these functions appears when one does not consider comparing two operations, but one operation and some given inputs and outputs. Suppose we have a user willing to provide his home town name and zip code in order to get a weather report. We could obviously look only for operations with both similar inputs and outputs, but this similarity function is most of the time too strict to produce relevant results, which is why we discarded it. It is the case here, since no operation corresponds exactly to the request. In this operation discovery context, the output the user desires is generally considered as the most important constraint (Keller, Lara, Lausen, Polleres & Fensel, 2005). FullSim can therefore be considered as the second best solution, since it includes all the desired outputs and a part of the available inputs. In our example, both $o_4$ and $o_5$ would be selected.

Still, it is possible to find no operation meeting these criteria, in which case the user might have to relax the constraints regarding his goal. In our example, suppose $o_4$ and $o_5$ are unavailable, and the user switches to ExcessSim. Operation $o_6$ is then the only solution, and it returns an additional weather report subscription compared to the initial request. It is likely the user will not be interested in this result, since he is looking for a free service. But on the contrary, he might have been interested in other additional outputs such as a list of weather reports for the neighboring cities. The PartialSim function works likewise, except it returns operations providing only a part of the desired goals (instead of more than the desired goals for ExcessSim).

The RelationSim function corresponds to a further relaxation of the user's constraints. Suppose the user is still looking for his weather report, but can only provide a zip code. If $o_4$ is unavailable, then no operation can be found using FullSim, PartialSim or ExcessSim. On the contrary, RelationSim will return $o_5$, which has the appropriate output but completely different input. Even if the operation cannot be invoked directly, further search might provide the needed parameters, for instance by mining a composition of operations. In our case the user could invoke $o_1$ first and take advantage of its output (a city name) to invoke $o_5$.

To conclude this section, we want to highlight the fact the proposed similarity functions were designed to be complementary, and not to be opposed. Each one corresponds to a specific use, directly related to the user goal. This is why, for completeness purpose, all four similarity networks will be build and analyzed in the next section.

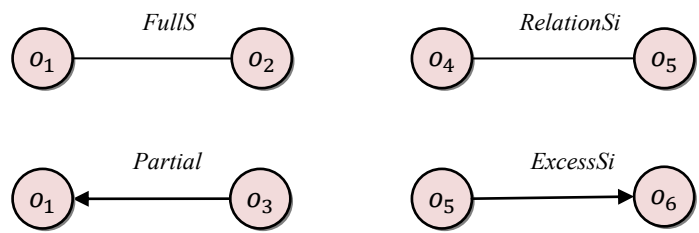

**Figure 1** *Similarity between WS operations*

## 5  Results and Discussion

We extracted similarity networks from the SAWSDL-TC1 collection of WS descriptions (Klusch & Kapahnke, 2008). This test collection provides 894 WS descriptions written in SAWSDL, and distributed over 7 thematic domains (education, medical care, food, travel, communication, economy and weapon). It originates in the OWLS-TC2.2 collection, which contains real-world WS descriptions retrieved from public IBM UDDI registries, and semi-automatically transformed from WSDL to OWL-S. This collection was subsequently resampled to increase its size, and converted to SAWSDL. From a SAWSDL file, we can extract the information needed to build our similarity networks.

**Network Structure and Components**

Each network contains 785 nodes, corresponding to the 785 operations of the collection. The first four rows in Table 2 summarize the major results we processed regarding the networks structure. Except for the first row, all the others properties are computed on the trimmed networks, i.e. without any isolated nodes. For all our networks, and unlike most real-world networks, no giant component is emerging, but numerous small ones, and isolated nodes. An example of this common structure is shown in Figure 2. This reflects the decomposition of the collection into a reasonable number of communities. This is a good thing, because having only isolated nodes or a giant component would lead to useless communities. Indeed, in the former case, each community would contain only one operation, and in the latter all operations would be considered as similar to the all others. Both cases would have been surprising considering we processed a real-world collection.

The number of isolated nodes globally decreases when going from FullSim to RelationSim, and at the same time, the numbers of links and components increase. Indeed, as constraints on outputs become less strict, more links are created leading to new components or increase of the existing ones. The numbers of operations, links and components are the highest in the RelationSim network. It means in this collection, a lot of operations produce identical outputs with completely different inputs. This contributes to increase the number of available and potentially usable operations in a discovery process.

Each component corresponds to a group of similar operations representing a community. For instance, in this PartialSim network component, operations `get_DESTINATION_HOTEL`, `get_SPORTS_HOTEL`, `get_ACTIVITY_HOTEL` are linked with `get_HOTEL`. Indeed `get_HOTEL` operations provides only the `HOTEL` output parameter while the three others provide the `HOTEL` output parameter and an additional specific one. A `get_HOTEL` operation can satisfy a destination/hotel request, an activity/hotel request or a sports/hotel request but not completely. In the RelationSim network, one component gathers operations that produce an output parameter named `get_LUXURYHOTEL`. It contains five operations. One of them has the parameter `CITY` as input, another one has the parameter `GEOGRAPHICAL-REGION` as input.

More than 90% of nodes and links are contained in the first 17, 30, 40, 32 communities in the FullSim, PartialSim, ExcessSim and RelationSim networks respectively. Table 3 shows range values for those principal communities. The most remarkable point is the links number of the largest RelationSim community. This corresponds to the PRICE output parameter which is common to several domains and sub-domains (food, car, book, device, economy). This number then falls to 149 for the RECOMMENDEDPRICE output parameter, and to 89 for the FUNDING output parameter, which are also shared by several domains.

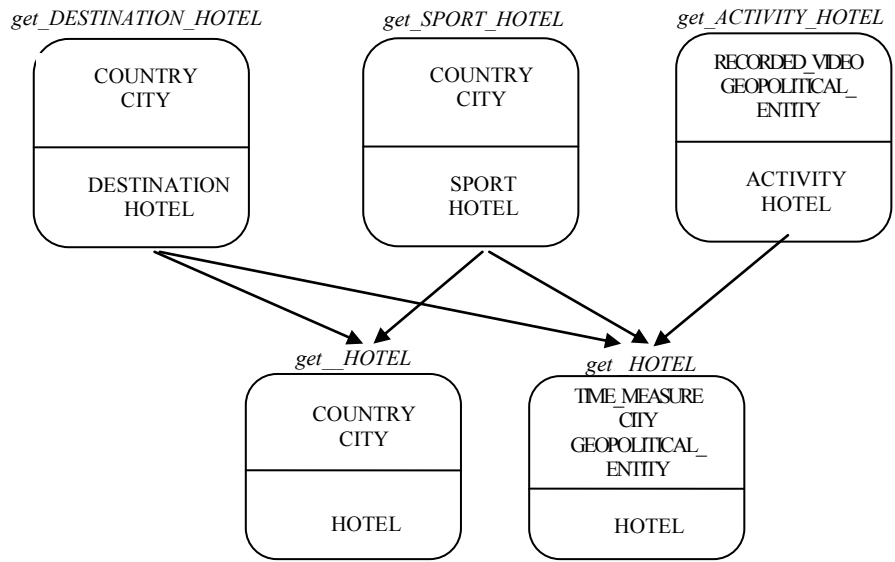

**Figure 2** *A PartialSim Component*

**Table 2** *Networks Properties*

| Property | FullSim | PartialSim | ExcessSim | RelationSim |
| --- | --- | --- | --- | --- |
| Isolated nodes | 604 | 447 | 486 | 227 |
| Nodes in trimmed network | 181 | 338 | 299 | 548 |
| Components | 38 | 61 | 67 | 123 |
| Links | 310 | 412 | 307 | 2254 |
| Average distance | 1.75 | 2.49 | 2.10 | 1.11 |
| Transitivity | 0.72 | 0.02 | 0.04 | 0.93 |

**Table 3** *Networks Components Properties*

| Property | FullSim | PartialSim | ExcessSim | RelationSim |
| --- | --- | --- | --- | --- |
| Nodes | 17 − 4 | 33 − 3 | 19 − 3 | 49 − 5 |
| Links | 53 − 4 | 74 − 2 | 38 − 2 | 1117 − 9 |

**Average Distance and Transitivity**

As shown in Table 2, all the similarity networks exhibit a small average distance. By comparison, this distance is approximately ranging from 3 to 10 in ER random networks of comparable sizes, which means the similarity networks possess the small-world property. In other words, many shortcuts exist in the networks, indicating that several operations share the same input parameters for FullSim, PartialSim and ExcessSim. For the RelationSim case, this observation confirms the previous point according to which a lot of operations produce same parameters with completely different inputs. Nevertheless, the fact small components are numerous also has an impact on this result, since non-existing paths are not taken into account when processing the average distance.

For all the similarity networks, the measured transitivity is higher than for comparable ER networks (whose transitivity is less than 0.01). It is not enough to conclude our networks have a high transitivity relatively to other real-world networks, though. However, given our networks do not have a giant component, this still means a significant number of triangles exist inside the communities. Consider for example the component containing the `get_SKILLEDOCCUPATION` operations in the FullSim network. The four operations have the `SKILLEDOCCUPATION` parameter as output. They are linked with four triangles because of their input parameters sets `{CITY, COUNTRY}`, `{COUNTRY}`, `{COUNTRY, PUBLICCOMPANY}` and `{COUNTRY, COMPANY}`.

# 6 Conclusion

In this paper, we proposed a new method for building WS communities, aiming at grouping WS operations which are similar in terms of functionalities. We designed four similarity functions, each one corresponding to a different definition of the concept of similarity. We described a method using a similarity function to build a similarity network of WS operations. In such a network, communities are topologically defined under the form of components, i.e. maximal connected subgraphs. As an example, we applied each function on a given real-world collection of WS to generate different similarity networks of operations. We shown our method allows identifying consistent communities of operations from these networks. We additionally discussed and compared the topological properties of the networks through the use of complex networks tools. All four networks exhibit a small average distance, which is a property observed in most real-world networks. At the opposite side, the transitivity property is small when compared to other real-world networks, but still high enough if we consider the networks do not exhibit any giant component.

The originality of our work lies in the similarity functions and in the similarity network-based method, which, to our knowledge, were never used in the context of WS communities building before. We additionally gave a analysis of the networks. We plan to extend it in two ways. First, the collection we used to build similarity networks is based on a set of real-world WS descriptions, but half of them were generated through resampling. Hence, it cannot be considered as perfectly realistic. We want to analyze similarity networks extract from a collection of real WS descriptions like the one found in (Hess, Johnston & Kushmerick, 2004) which contain 800 WSDL files. Second, it would be interesting to build networks based on semantic descriptions and compare their properties to those of the syntactic networks presented in this article. As semantics seems to improve WS discovery and composition, we can expect better results for semantic network-based communities.